# Multiple-Input Multiple-Output Gaussian Broadcast Channels with Common and Confidential Messages

Hung D. Ly, Tie Liu, and Yingbin Liang[*]

October 29, 2018

## Abstract

This paper considers the problem of the multiple-input multiple-output (MIMO) Gaussian broadcast channel with two receivers (receivers 1 and 2) and two messages: a common message intended for both receivers and a confidential message intended only for receiver 1 but needing to be kept asymptotically perfectly secure from receiver 2. A matrix characterization of the secrecy capacity region is established via a channel enhancement argument. The enhanced channel is constructed by first splitting receiver 1 into two virtual receivers and then enhancing only the virtual receiver that decodes the confidential message. The secrecy capacity region of the enhanced channel is characterized using an extremal entropy inequality previously established for characterizing the capacity region of a degraded compound MIMO Gaussian broadcast channel.

## 1 Introduction

Understanding the fundamental limits of multiple-input multiple-output (MIMO) secrecy communication is an important research topic in wireless physical layer security. A basic model of MIMO secrecy communication is a MIMO Gaussian broadcast channel with two receivers, for which the channel outputs at time index $m$ are given by

$$\mathbf{Y}_k[m] = \mathbf{H}_k \mathbf{X}[m] + \mathbf{Z}_k[m], \quad k = 1, 2 \tag{1}$$

where $\mathbf{H}_k$ is the (real) channel matrix of size $r_k \times t$ for receiver $k$, and $\{\mathbf{Z}_k[m]\}_m$ is an independent and identically distributed (i.i.d.) additive vector Gaussian noise process with

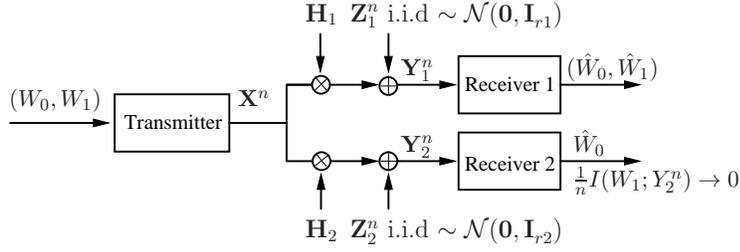

(a) The general case

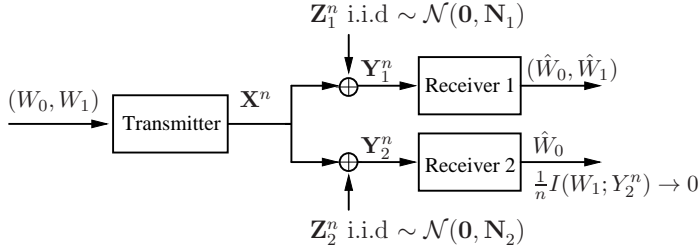

(b) The aligned case

Figure 1: MIMO Gaussian broadcast channel with common and confidential messages.

zero mean and identity covariance matrix. The channel input $\{\mathbf{X}[m]\}_m$ is subject to an average total power constraint:
$$\frac{1}{n}\sum_{m=1}^{n}\|\mathbf{X}[m]\|^2 \leq P. \tag{2}$$

The transmitter has a set of two independent messages $(W_0, W_1)$, where $W_0$ is a common message intended for both receivers 1 and 2, and $W_1$ is a confidential message intended for receiver 1 but needing to be kept secret from receiver 2. The confidentiality of message $W_1$ at receiver 2 is measured using the information-theoretic criterion [1, 2]:
$$\frac{1}{n}I(W_1; \mathbf{Y}_2^n) \to 0 \tag{3}$$
where $\mathbf{Y}_2^n := (\mathbf{Y}_2[1], \ldots, \mathbf{Y}_2[n])$, and the limit is taken as the blocklength $n \to \infty$. An illustration of this communication scenario is shown in Figure 1(a). The goal is to characterize the entire rate region $\mathcal{C}_s(\mathbf{H}_1, \mathbf{H}_2, P)$ that includes all rate pairs $(R_0, R_1)$ that can be achieved by any coding scheme. In this paper, we term $\mathcal{C}_s(\mathbf{H}_1, \mathbf{H}_2, P)$ as the *secrecy capacity region* despite the fact that $W_0$ is not a confidential message.

In their seminar work [2], Csiszár and Körner considered the discrete memoryless case of the problem. A single-letter expression of the secrecy capacity region was given as the set of nonnegative rate pairs $(R_0, R_1)$ satisfying
$$\begin{aligned} R_0 &\leq \min\left[I(U;Y_1), I(U;Y_2)\right] \\ R_1 &\leq I(V;Y_1|U) - I(V;Y_2|U) \end{aligned} \tag{4}$$

for some $p(u,v,x,y_1,y_2) = p(u)p(v|u)p(x|v)p(y_1,y_2|x)$, where $p(y_1,y_2|x)$ is the transition probability of the discrete memoryless broadcast channel. Thus, in principle, the secrecy capacity region $\mathcal{C}_s(\mathbf{H}_1,\mathbf{H}_2,P)$ can be computed by evaluating the Csiszár-Körner region (4) for the MIMO Gaussian broadcast channel (1).

However, directly evaluating (4) for the MIMO Gaussian broadcast channel (1) appears difficult due to the presence of the auxiliary variables $U$ and $V$. Consider, for example, the special case where the common message $W_0$ is absent, i.e., $R_0 = 0$. Let $U$ be deterministic in (4). Then, the maximum of $R_1$ can be determined by solving the optimization program

$$\max_{p(x,v)} \left[ I(V;Y_1) - I(V;Y_2) \right]. \tag{5}$$

In literature, the problem of communicating a confidential message over a MIMO Gaussian broadcast channel is termed as the MIMO Gaussian wiretap channel problem. Characterizing the secrecy capacity of the MIMO Gaussian wiretap channel has been an active area of research in recent years. However, despite intensive effort [3, 4, 5, 6, 7], determining the secrecy capacity of the MIMO Gaussian wiretap channel via *directly* solving the optimization program (5) remains intractable.

Recently, Khisti and Wornell [5] and Oggier and Hassibi [6] studied the MIMO Gaussian wiretap channel problem and proposed an *indirect* approach to solve the optimization program (5). The main idea was to compute an upper bound on the secrecy capacity by considering a fictitious MIMO Gaussian wiretap channel in which the legitimate receiver has access to both received signals $\{\mathbf{Y}_1[m]\}_m$ and $\{\mathbf{Y}_2[m]\}_m$. For any fixed correlation between the additive noise $\mathbf{Z}_1[m]$ and $\mathbf{Z}_2[m]$, Khisti and Wornell [5] and Oggier and Hassibi [6] showed that *Gaussian* random binning *without* prefix coding is optimal for the fictitious channel. Comparing the upper bound (minimized over all possible correlations between $\mathbf{Z}_1[m]$ and $\mathbf{Z}_2[m]$) with the achievable secrecy rate by choosing a *Gaussian* $V = \mathbf{X}$ in the objective function of (5) established an exact matrix characterization of the secrecy capacity. However, matching the upper and lower bounds requires complicated matrix analysis, which makes the approach difficult to extend to the more general scenario with both common and confidential messages.

More recently, Liu and Shamai [7] presented an alternative, simpler characterization of the secrecy capacity of the MIMO Gaussian wiretap channel. Compared with the work of [5] and [6], there are two key differences in the argument of [7]

1. Instead of the average total power constraint (2), [7] considered the more general matrix power constraint:

$$\frac{1}{n} \sum_{m=1}^{n} \left( \mathbf{X}[m]\mathbf{X}^{\mathrm{T}}[m] \right) \preceq \mathbf{S} \tag{6}$$

   where $\mathbf{S}$ is a positive semidefinite matrix, and "$\preceq$" denotes "less than or equal to" in the positive semidefinite ordering between real symmetric matrices.

2. Different from the Sato-like [8] argument of [5] and [6], the upper bound on the secrecy capacity in [7] was obtained by considering an *enhanced* MIMO Gaussian wiretap channel that has the *same* secrecy capacity as the original wiretap channel. Channel-enhancement argument was first introduced by Weingarten et al. [9] to characterize the *private message* capacity region of the MIMO Gaussian broadcast channel; [7] was the first to apply this argument to MIMO *secrecy* communication problems.

The main goal of this paper is to *adapt* the channel-enhancement argument of [7] to the more general problem of MIMO Gaussian broadcast channel with both common and confidential messages. Our main result is that for the MIMO Gaussian broadcast channel (1), a jointly *Gaussian* $(U, V, \mathbf{X})$ with $V = \mathbf{X}$ is *optimal* for the Csiszár-Körner region (4). This establishes a matrix characterization of the secrecy capacity region of the MIMO Gaussian broadcast channel under a matrix power constraint.

The rest of the paper is organized as follows. In Section 2, we summarize the main results of the paper. In Section 3, we consider the special case of the MIMO Gaussian broadcast channel (1) in which the channel matrices $\mathbf{H}_1$ and $\mathbf{H}_2$ are square and invertible, and use a channel-enhancement argument to characterize the secrecy capacity region. In Section 4, we broaden the result of Section 3, via a limiting argument, to the general case, and characterize the secrecy capacity region of the general MIMO Gaussian broadcast channel. Finally, in Section 5, we provide some numerical examples to illustrate the main results of the paper.

## 2 Main Results

The following theorem summarizes the secrecy capacity region of the MIMO Gaussian broadcast channel with common and confidential messages under a matrix power constraint.

**Theorem 1.** *The secrecy capacity region $\mathcal{C}_s(\mathbf{H}_1, \mathbf{H}_2, \mathbf{S})$ of the MIMO Gaussian broadcast channel* (1) *with messages $W_0$ (intended for both receivers 1 and 2) and $W_1$ (intended for receiver 1 but needing to be kept asymptotically perfectly secret from receiver 2) under the matrix power constraint* (6) *is given by the set of all nonnegative rate pairs $(R_0, R_1)$ satisfying*

$$
\begin{aligned}
R_0 &\leq \min\left(\tfrac{1}{2}\log\left|\tfrac{\mathbf{H}_1\mathbf{S}\mathbf{H}_1^\mathrm{T}+\mathbf{I}_{r_1}}{\mathbf{H}_1\mathbf{B}\mathbf{H}_1^\mathrm{T}+\mathbf{I}_{r_1}}\right|, \tfrac{1}{2}\log\left|\tfrac{\mathbf{H}_2\mathbf{S}\mathbf{H}_2^\mathrm{T}+\mathbf{I}_{r_2}}{\mathbf{H}_2\mathbf{B}\mathbf{H}_2^\mathrm{T}+\mathbf{I}_{r_2}}\right|\right) \\
R_1 &\leq \tfrac{1}{2}\log\left|\mathbf{H}_1\mathbf{B}\mathbf{H}_1^\mathrm{T}+\mathbf{I}_{r_1}\right| - \tfrac{1}{2}\log\left|\mathbf{H}_2\mathbf{B}\mathbf{H}_2^\mathrm{T}+\mathbf{I}_{r_2}\right|
\end{aligned}
\qquad(7)
$$

*for some $0 \preceq \mathbf{B} \preceq \mathbf{S}$. Here, $\mathbf{I}_{r_k}$ denotes the identity matrix of size $r_k \times r_k$.*

As mentioned previously, the MIMO Gaussian wiretap channel problem can be considered as a special case here with the common rate $R_0 = 0$. We have thus recovered the main result of [7], restated below as a corollary.

**Corollary 2** ([7]). *The secrecy capacity $C_s(\mathbf{H}_1, \mathbf{H}_2, \mathbf{S})$ of the MIMO Gaussian broadcast channel* (1) *with a confidential message messages $W$ (intended for receiver 1 but needing to be kept asymptotically perfectly secret from receiver 2) under the matrix power constraint* (6) *is given by*

$$
C_s(\mathbf{H}_1, \mathbf{H}_2, \mathbf{S}) = \max_{0 \preceq \mathbf{B} \preceq \mathbf{S}} \left(\frac{1}{2}\log\left|\mathbf{H}_1\mathbf{B}\mathbf{H}_1^\mathrm{T} + \mathbf{I}_{r_1}\right| - \frac{1}{2}\log\left|\mathbf{H}_2\mathbf{B}\mathbf{H}_2^\mathrm{T} + \mathbf{I}_{r_2}\right|\right). \qquad(8)
$$

In engineering practice, it is particularly relevant to consider the average total power constraint. The following corollary summarizes the secrecy capacity region of the MIMO Gaussian broadcast channel with common and confidential messages under an average total power constraint. The result is a simple consequence of [9, Lemma 1].

**Corollary 3.** *The secrecy capacity region $\mathcal{C}_s(\mathbf{H}_1, \mathbf{H}_2, P)$ of the MIMO Gaussian broadcast channel* (1) *with messages $W_0$ (intended for both receivers 1 and 2) and $W_1$ (intended for receiver 1 but needing to be kept asymptotically perfectly secret from receiver 2) under the average total power constraint* (2) *is given by the set of all nonnegative rate pairs $(R_0, R_1)$ satisfying*

$$\begin{aligned} R_0 &\leq \min\left[\tfrac{1}{2}\log\left|\tfrac{\mathbf{H}_1(\mathbf{B}_1+\mathbf{B}_2)\mathbf{H}_1^T+\mathbf{I}_{r_1}}{\mathbf{H}_1\mathbf{B}_1\mathbf{H}_1^T+\mathbf{I}_{r_1}}\right|, \tfrac{1}{2}\log\left|\tfrac{\mathbf{H}_2(\mathbf{B}_1+\mathbf{B}_2)\mathbf{H}_2^T+\mathbf{I}_{r_2}}{\mathbf{H}_2\mathbf{B}_1\mathbf{H}_2^T+\mathbf{I}_{r_2}}\right|\right] \\ R_1 &\leq \tfrac{1}{2}\log\left|\mathbf{H}_1\mathbf{B}_1\mathbf{H}_1^T+\mathbf{I}_{r_1}\right| - \tfrac{1}{2}\log\left|\mathbf{H}_2\mathbf{B}_1\mathbf{H}_2^T+\mathbf{I}_{r_2}\right| \end{aligned} \quad (9)$$

*for some positive semidefinite matrices $\mathbf{B}_1$ and $\mathbf{B}_2$ with $\mathrm{Tr}(\mathbf{B}_1+\mathbf{B}_2) \leq P$.*

The achievability proof of Theorem 1 follows from the Csiszár-Körner region (4) by letting $U$ be a $t$-dimensional Gaussian vector with zero mean and covariance matrix $\mathbf{S} - \mathbf{B}$ and $V = \mathbf{X} = U + G$, where $G$ is a $t$-dimensional Gaussian vector with zero mean and covariance matrix $\mathbf{B}$ and is independent of $U$. Note that prefix coding is *not* needed in communicating the confidential message $W_1$ even though the corresponding eavesdropper channel may not be degraded with respect to the legitimate receiver channel.

The converse of Theorem 1 follows from an adaptation of the channel-enhancement argument of [7] with the following two new ingredients:

1. To obtain an enhanced MIMO Gaussian broadcast channel that has the *same* weighted secrecy sum-capacity as the original channel, we need to split receiver 1 into two *virtual* receivers: one as the legitimate receiver for the confidential message $W_1$, and the other as one of the intended receivers for the common message $W_0$. Only the legitimate receiver for the confidential message $W_1$ is enhanced in the proof.

2. With only a confidential message, in [7], the matrix characterization of the secrecy capacity of the enhanced channel was obtained via the worst noise result of Diggavi and Cover [13]. With both common and confidential messages, characterizing the secrecy capacity region of the enhanced channel becomes more involved. In our proof, we resort to an extremal entropy inequality which was first proved by Weingarten et al. [12] for characterizing the capacity region of a degraded *compound* MIMO Gaussian broadcast channel.

The details of the proof are provided in the next two sections.

## 3 Aligned MIMO Gaussian Broadcast Channel

In this section, we prove Theorem 1 for the special case where the channel matrices $\mathbf{H}_1$ and $\mathbf{H}_2$ are square and invertible. In this case, multiplying both sides of (1) by $\mathbf{H}_k^{-1}$, the channel model can be equivalently written as

$$\mathbf{Y}_k[m] = \mathbf{X}[m] + \mathbf{Z}_k[m], \qquad k = 1, 2 \tag{10}$$

where $\{\mathbf{Z}_k[m]\}_m$ is an i.i.d. additive vector Gaussian noise process with zero mean and covariance matrix

$$\mathbf{N}_k = \mathbf{H}_k^{-1} \mathbf{H}_k^{-\mathrm{T}}.$$

Following [9], we will term the channel model (10) as the *aligned* MIMO Gaussian broadcast channel (see Figure 1(b)) and (1) as the *general* MIMO Gaussian broadcast channel. The main result of this section is summarized in the following theorem.

**Theorem 4.** *The secrecy capacity region $\mathcal{C}_s(\mathbf{N}_1, \mathbf{N}_2, \mathbf{S})$ of the aligned MIMO Gaussian broadcast channel (10) with messages $W_0$ (intended for both receivers 1 and 2) and $W_1$ (intended for receiver 1 but needing to be kept asymptotically perfectly secret from receiver 2) under the matrix power constraint (6) is given by the set of all nonnegative rate pairs $(R_0, R_1)$ satisfying*

$$\begin{aligned} R_0 &\leq \min\left(\tfrac{1}{2}\log\left|\tfrac{\mathbf{S}+\mathbf{N}_1}{\mathbf{B}+\mathbf{N}_1}\right|, \tfrac{1}{2}\log\left|\tfrac{\mathbf{S}+\mathbf{N}_2}{\mathbf{B}+\mathbf{N}_2}\right|\right) \\ R_1 &\leq \tfrac{1}{2}\log\left|\tfrac{\mathbf{B}+\mathbf{N}_1}{\mathbf{N}_1}\right| - \tfrac{1}{2}\log\left|\tfrac{\mathbf{B}+\mathbf{N}_2}{\mathbf{N}_2}\right| \end{aligned} \tag{11}$$

*for some $0 \preceq \mathbf{B} \preceq \mathbf{S}$.*

*Proof.* Let $G$ be a $t$-dimensional Gaussian vector with zero mean and covariance matrix $\mathbf{B}$. Then, the achievability of (11) can be obtained from the Csiszár-Körner region (4) by letting $U$ be a $t$-dimensional Gaussian vector with zero mean and covariance matrix $\mathbf{S}-\mathbf{B}$ and $V = \mathbf{X} = U + G$, where $U$ and $G$ are assumed to be independent. We therefore concentrate on proving the converse result.

To show that any achievable secrecy rate pair $(R_0, R_1)$ for the aligned MIMO Gaussian broadcast channel (10) must satisfy (11) for some $0 \preceq \mathbf{B} \preceq \mathbf{S}$, we may assume, without loss of generality, that the matrix power constraint $\mathbf{S} \succ 0$. For the case where $\mathbf{S} \succeq 0$ but $|\mathbf{S}| = 0$, let $\theta = \mathsf{Rank}(\mathbf{S}) < t$. We can define an *equivalent* aligned MIMO Gaussian broadcast channel with $\theta$ transmit and receive antennas and a new covariance matrix power constraint that is strictly positive definite. Hence, we can convert the case where $\mathbf{S} \succeq 0$, $|\mathbf{S}| = 0$ to the case where $\mathbf{S} \succ 0$ with the same secrecy capacity region. See [9, Lemma 2] for a formal presentation of this argument.

For the case where $\mathbf{S} \succ 0$, we shall consider proof by contradiction as follows. Assume that $(R_0^o, R_1^o)$ is an achievable secrecy rate pair for the aligned MIMO Gaussian broadcast channel (10) that lies *outside* the rate region (11). Since $(R_0^o, R_1^o)$ is achievable, $R_0^o$ can be bounded from above as

$$R_0^o \leq \min\left(\frac{1}{2}\log\left|\frac{\mathbf{S}+\mathbf{N}_1}{\mathbf{N}_1}\right|, \frac{1}{2}\log\left|\frac{\mathbf{S}+\mathbf{N}_2}{\mathbf{N}_2}\right|\right) = R_0^{max}. \tag{12}$$

Moreover, if $R_1^o = 0$, then $R_0^{max}$ can be achieved by letting $\mathbf{B} = 0$ in (11). Therefore, we can write $R_1^o = R_1^* + \delta$ for some $\delta > 0$, where $R_1^*$ is given by

$$\begin{aligned}
\max_{\mathbf{B}} \quad & \tfrac{1}{2}\log\left|\tfrac{\mathbf{B}+\mathbf{N}_1}{\mathbf{N}_1}\right| - \tfrac{1}{2}\log\left|\tfrac{\mathbf{B}+\mathbf{N}_2}{\mathbf{N}_2}\right| \\
\text{subject to} \quad & \tfrac{1}{2}\log\left|\tfrac{\mathbf{S}+\mathbf{N}_1}{\mathbf{B}+\mathbf{N}_1}\right| \geq R_0^o \\
& \tfrac{1}{2}\log\left|\tfrac{\mathbf{S}+\mathbf{N}_2}{\mathbf{B}+\mathbf{N}_2}\right| \geq R_0^o \\
& 0 \preceq \mathbf{B} \preceq \mathbf{S}.
\end{aligned} \quad (13)$$

The above optimization program can be rewritten in the following standard form:

$$\begin{aligned}
\min_{\mathbf{B}} \quad & \tfrac{1}{2}\log\left|\tfrac{\mathbf{B}+\mathbf{N}_2}{\mathbf{N}_2}\right| - \tfrac{1}{2}\log\left|\tfrac{\mathbf{B}+\mathbf{N}_1}{\mathbf{N}_1}\right| \\
\text{subject to} \quad & R_0^o - \tfrac{1}{2}\log\left|\tfrac{\mathbf{S}+\mathbf{N}_1}{\mathbf{B}+\mathbf{N}_1}\right| \leq 0 \\
& R_0^o - \tfrac{1}{2}\log\left|\tfrac{\mathbf{S}+\mathbf{N}_2}{\mathbf{B}+\mathbf{N}_2}\right| \leq 0 \\
& -\mathbf{B} \preceq 0 \\
& \mathbf{B} - \mathbf{S} \preceq 0
\end{aligned} \quad (14)$$

which has one semidefinite variable, $\mathbf{B}$, constrained by both scalar and semidefinite inequalities. This is in fact an optimization problem with generalized constraints in the form of semidefinite inequalities [14, p. 267]. Therefore, the Karush-Kuhn-Tucker (KKT) condition states that the derivative of the Lagrangian

$$\mathcal{L} = \left(\tfrac{1}{2}\log\left|\tfrac{\mathbf{B}+\mathbf{N}_2}{\mathbf{N}_2}\right| - \tfrac{1}{2}\log\left|\tfrac{\mathbf{B}+\mathbf{N}_1}{\mathbf{N}_1}\right|\right) + \sum_{k=1}^{2}\mu_k\left(R_0^o - \tfrac{1}{2}\log\left|\tfrac{\mathbf{S}+\mathbf{N}_k}{\mathbf{B}+\mathbf{N}_k}\right|\right) + \\ \mathrm{Tr}\left((-\mathbf{B})\mathbf{M}_1\right) + \mathrm{Tr}\left((\mathbf{B}-\mathbf{S})\mathbf{M}_2\right) \quad (15)$$

must vanish at an optimal solution $\mathbf{B}^*$.[1] Here, $\mathbf{M}_k$, $k = 1, 2$, are positive semidefinite matrices such that

$$\mathbf{B}^*\mathbf{M}_1 = 0 \quad (16)$$
$$(\mathbf{S} - \mathbf{B}^*)\mathbf{M}_2 = 0 \quad (17)$$

and $\mu_k \geq 0$, $k = 1, 2$, with equality if

$$\frac{1}{2}\log\left|\frac{\mathbf{S}+\mathbf{N}_k}{\mathbf{B}^*+\mathbf{N}_k}\right| > R_0^o.$$

We immediately have

$$\mu_k R_0^o = \frac{\mu_k}{2}\log\left|\frac{\mathbf{S}+\mathbf{N}_k}{\mathbf{B}^*+\mathbf{N}_k}\right|, \quad k = 1, 2. \quad (18)$$

Taking derivative of the Lagrangian in (15) over $\mathbf{B}$, the KKT condition can be written as

$$\nabla_{\mathbf{B}}\left(\frac{1}{2}\log\left|\frac{\mathbf{B}+\mathbf{N}_2}{\mathbf{N}_2}\right| - \frac{1}{2}\log\left|\frac{\mathbf{B}+\mathbf{N}_1}{\mathbf{N}_1}\right| + \sum_{k=1}^{2}\mu_k\left(R_0^o - \frac{1}{2}\log\left|\frac{\mathbf{S}+\mathbf{N}_k}{\mathbf{B}+\mathbf{N}_k}\right|\right)\right) - \mathbf{M}_1 + \mathbf{M}_2 = 0$$

---

[1] As this optimization problem is not necessarily convex, a set of constraint qualifications (CQs) should be verified to make sure that the KKT conditions indeed hold. The CQs stated in [9, Appendix D] hold in a trivial manner for this optimization program.

which gives
$$\frac{1}{2}(\mathbf{B}^* + \mathbf{N}_1)^{-1} + \mathbf{M}_1 = \frac{\mu_1}{2}(\mathbf{B}^* + \mathbf{N}_1)^{-1} + \frac{\mu_2 + 1}{2}(\mathbf{B}^* + \mathbf{N}_2)^{-1} + \mathbf{M}_2. \quad (19)$$

By (13) and (18), we have
$$R_1^o + (\mu_1 + \mu_2)R_0^o = \frac{1}{2}\log\left|\frac{\mathbf{B}^* + \mathbf{N}_1}{\mathbf{N}_1}\right| - \frac{1}{2}\log\left|\frac{\mathbf{B}^* + \mathbf{N}_2}{\mathbf{N}_2}\right| + \sum_{k=1}^{2}\left(\frac{\mu_k}{2}\log\left|\frac{\mathbf{S} + \mathbf{N}_k}{\mathbf{B}^* + \mathbf{N}_k}\right|\right) + \delta. \quad (20)$$

Next, we shall find a contradiction to (20) by showing that for *any* achievable secrecy rate pair $(R_0, R_1)$,
$$R_1 + (\mu_1 + \mu_2)R_0 \leq \frac{1}{2}\log\left|\frac{\mathbf{B}^* + \mathbf{N}_1}{\mathbf{N}_1}\right| - \frac{1}{2}\log\left|\frac{\mathbf{B}^* + \mathbf{N}_2}{\mathbf{N}_2}\right| + \sum_{k=1}^{2}\left(\frac{\mu_k}{2}\log\left|\frac{\mathbf{S} + \mathbf{N}_k}{\mathbf{B}^* + \mathbf{N}_k}\right|\right).$$

We divide our proof into three steps.

**Step 1: Split receiver 1 into two virtual receivers.**

Consider the following aligned MIMO Gaussian broadcast channel with three receivers:
$$\begin{aligned} \mathbf{Y}_{1a}[m] &= \mathbf{X}[m] + \mathbf{Z}_{1a}[m] \\ \mathbf{Y}_{1b}[m] &= \mathbf{X}[m] + \mathbf{Z}_{1b}[m] \\ \mathbf{Y}_2[m] &= \mathbf{X}[m] + \mathbf{Z}_2[m] \end{aligned} \quad (21)$$

where $\{\mathbf{Z}_{1a}[m]\}_m$, $\{\mathbf{Z}_{1b}[m]\}_m$ and $\{\mathbf{Z}_2[m]\}_m$ are i.i.d. additive vector Gaussian noise processes with zero means and covariance matrices $\mathbf{N}_1$, $\mathbf{N}_1$ and $\mathbf{N}_2$, respectively. Suppose that the transmitter has two independent messages $W_0$ and $W_1$, where $W_0$ is intended for both receivers 1b and 2 and $W_1$ is intended for receiver 1a but needs to be kept secret from receiver 2. The confidentiality of message $W_1$ at receiver 2 is measured using the information-theoretic criterion (3). See Figure 2(a) for an illustration of this communication scenario.

Note that both receivers 1a and 1b in the aligned MIMO Gaussian broadcast channel (21) have the same noise covariance matrices as receiver 1 in the aligned MIMO Gaussian broadcast channel (10), and receiver 2 in the aligned MIMO Gaussian broadcast channel (21) has the same noise covariance matrix as receiver 2 in the aligned MIMO Gaussian broadcast channel (10). Therefore, any achievable secrecy rate pair $(R_0, R_1)$ for the aligned MIMO Gaussian broadcast channel (21) can also be achieved by the *same* coding scheme for the aligned MIMO Gaussian broadcast channel (10), and vice versa. Thus, the aligned MIMO Gaussian broadcast channel (21) has the *same* secrecy capacity region as the aligned MIMO Gaussian broadcast channel in (10) under the same power constraints.

**Step 2: Construct an enhanced channel.**

Let $\tilde{\mathbf{N}}_1$ be a real symmetric matrix satisfying
$$\frac{1}{2}(\mathbf{B}^* + \tilde{\mathbf{N}}_1)^{-1} = \frac{1}{2}(\mathbf{B}^* + \mathbf{N}_1)^{-1} + \mathbf{M}_1. \quad (22)$$

Following [9, Lemma 11], we have
$$0 \prec \tilde{\mathbf{N}}_1 \preceq \mathbf{N}_1 \quad (23)$$

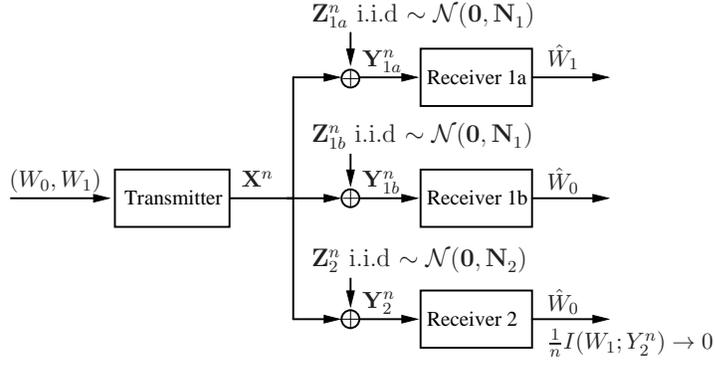

(a) An equivalent view of the aligned MIMO Gaussian broadcast channel shown in Figure 1(b)

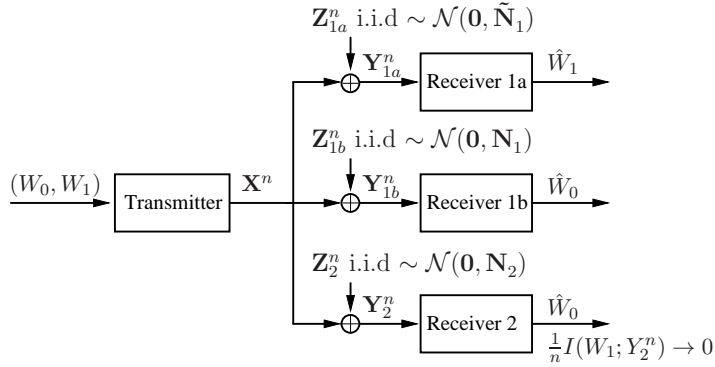

(b) The enhanced channel

Figure 2: Enhanced MIMO Gaussian broadcast channel with common and confidential messages.

and
$$\left|\frac{\mathbf{B}^* + \tilde{\mathbf{N}}_1}{\tilde{\mathbf{N}}_1}\right| = \left|\frac{\mathbf{B}^* + \mathbf{N}_1}{\mathbf{N}_1}\right|. \tag{24}$$

Moreover, substitute (22) into (19) and we have
$$\frac{1}{2}(\mathbf{B}^* + \tilde{\mathbf{N}}_1)^{-1} = \frac{\mu_1}{2}(\mathbf{B}^* + \mathbf{N}_1)^{-1} + \frac{\mu_2 + 1}{2}(\mathbf{B}^* + \mathbf{N}_2)^{-1} + \mathbf{M}_2. \tag{25}$$

Note that $(\mathbf{B}^* + \mathbf{N}_1)^{-1}$, $(\mathbf{B}^* + \mathbf{N}_2)^{-1}$ and $\mathbf{M}_2$ are all positive semidefinite so we have
$$\frac{1}{2}(\mathbf{B}^* + \tilde{\mathbf{N}}_1)^{-1} \succeq \frac{1}{2}(\mathbf{B}^* + \mathbf{N}_2)^{-1}$$

and hence
$$\tilde{\mathbf{N}}_1 \preceq \mathbf{N}_2. \tag{26}$$

Now consider the following enhanced MIMO Gaussian broadcast channel (see Figure 2(b)):

$$\begin{aligned} \tilde{\mathbf{Y}}_{1a}[m] &= \mathbf{X}[m] + \tilde{\mathbf{Z}}_{1a}[m] \\ \mathbf{Y}_{1b}[m] &= \mathbf{X}[m] + \mathbf{Z}_{1b}[m] \\ \mathbf{Y}_2[m] &= \mathbf{X}[m] + \mathbf{Z}_2[m] \end{aligned} \quad (27)$$

where $\{\tilde{\mathbf{Z}}_{1a}[m]\}_m$, $\{\mathbf{Z}_{1b}[m]\}_m$ and $\{\mathbf{Z}_2[m]\}_m$ are i.i.d. additive vector Gaussian noise processes with zero mean and covariance matrix $\tilde{\mathbf{N}}_1$, $\mathbf{N}_1$ and $\mathbf{N}_2$, respectively. Note from (23) that $\tilde{\mathbf{N}}_1 \preceq \mathbf{N}_1$. We conclude that the secrecy capacity region of the enhanced MIMO Gaussian broadcast channel (27) is at least as large as the secrecy capacity region of the aligned MIMO Gaussian broadcast channel (21) under the same power constraints.

**Step 3: Bound from above the weighted secrecy sum-capacity of the enhanced channel.**

Note from (23) and (26) that

$$0 \prec \tilde{\mathbf{N}}_1 \preceq \{\mathbf{N}_1, \mathbf{N}_2\}. \quad (28)$$

Thus, in the enhanced MIMO Gaussian broadcast channel (27), the received signals $\mathbf{Y}_{1b}[m]$ and $\mathbf{Y}_2[m]$ are (stochastically) degraded with respect to the received signal $\tilde{\mathbf{Y}}_{1a}[m]$. In the following proposition, we shall consider the discrete memoryless case of the enhanced channel (27) and provide a single-letter characterization of the secrecy capacity region.

**Proposition 1.** *Consider a discrete memoryless broadcast channel with transition probability $p(\tilde{y}_{1a}, y_{1b}, y_2|x)$ and messages $W_0$ (intended for both receivers 1b and 2) and $W_1$ (intended for receiver 1a but needs to be kept confidential from receiver 2). If both*

$$X \to \tilde{Y}_{1a} \to Y_{1b} \quad \text{and} \quad X \to \tilde{Y}_{1a} \to Y_2$$

*form Markov chains in their respective order, then the secrecy capacity region of this channel is given by the set of nonnegative rate pairs $(R_0, R_1)$ satisfying*

$$\begin{aligned} R_0 &\leq \min\left[I(U; Y_{1b}), I(U; Y_2)\right] \\ R_1 &\leq I(X; \tilde{Y}_{1a}|U) - I(X; Y_2|U) \end{aligned} \quad (29)$$

*for some $p(u, x, \tilde{y}_{1a}, y_{1b}, y_2) = p(u)p(x|u)p(\tilde{y}_{1a}, y_{1b}, y_2|x)$.*

*Proof.* The achievability of (29) follows from a coding scheme that combines superposition coding [11] and random binning [1]. The converse proof follows from the steps similar to those in the converse proof in [2]. The details of the converse proof are provided in the Appendix. □

**Remark 1.** *Prefix coding is no longer needed due to the preexisting Markov relation $X \to \tilde{Y}_{1a} \to Y_2$.*

Next, to evaluate the single-letter expression (29) for the enhanced MIMO Gaussian broadcast channel (27), we shall recall an extremal entropy inequality which is a special case of [12, Corollary 4].

**Proposition 2** ([12]). *Let $\tilde{\mathbf{Z}}_{1a}$, $\mathbf{Z}_{1b}$ and $\mathbf{Z}_2$ be t-dimensional Gaussian vectors with zero means and covariance matrices $\tilde{\mathbf{N}}_1$, $\mathbf{N}_1$ and $\mathbf{N}_2$, respectively. Assume that $\tilde{\mathbf{N}}_1$, $\mathbf{N}_1$ and $\mathbf{N}_2$ are ordered as in (28). Let $\mathbf{S}$ be a $t \times t$ positive definite matrix. If there exists a $t \times t$ real symmetric matrix $\mathbf{B}^*$ such that $0 \preceq \mathbf{B}^* \preceq \mathbf{S}$ and satisfying*

$$\tfrac{1}{2}(\mathbf{B}^* + \tilde{\mathbf{N}}_1)^{-1} = \tfrac{\mu\lambda}{2}(\mathbf{B}^* + \mathbf{N}_1)^{-1} + \tfrac{\mu(1-\lambda)}{2}(\mathbf{B}^* + \mathbf{N}_2)^{-1} + \mathbf{M}_2$$
$$(\mathbf{S} - \mathbf{B}^*)\mathbf{M}_2 = 0$$

*for some positive semidefinite matrix $\mathbf{M}_2$ and real scalars $\mu \geq 0$ and $0 \leq \lambda \leq 1$, then*

$$h(\mathbf{X}+\tilde{\mathbf{Z}}_{1a}|U) - \mu\lambda h(\mathbf{X}+\mathbf{Z}_{1b}|U) - \mu(1-\lambda)h(\mathbf{X}+\mathbf{Z}_2|U)$$
$$\leq \frac{1}{2}\log|2\pi e(\mathbf{B}^* + \tilde{\mathbf{N}}_1)| - \frac{\mu\lambda}{2}\log|2\pi e(\mathbf{B}^* + \mathbf{N}_1)| - \frac{\mu(1-\lambda)}{2}\log|2\pi e(\mathbf{B}^* + \mathbf{N}_2)|$$

*for any $(\mathbf{X}, U)$ independent of $(\tilde{\mathbf{Z}}_{1a}, \mathbf{Z}_{1b}, \mathbf{Z}_2)$ such that $E[\mathbf{X}\mathbf{X}^{\mathrm{T}}] \preceq \mathbf{S}$.*

We are now ready to bound from above the weighted secrecy sum-capacity of the enhanced channel (27). By Proposition 1, for any achievable secrecy rate pair $(R_0, R_1)$ for the enhanced channel (27) we have

$$\begin{aligned}
R_1 + (\mu_1 + \mu_2)R_0 &\leq I(\mathbf{X}; \tilde{\mathbf{Y}}_{1a}|U) - I(\mathbf{X}; \mathbf{Y}_2|U) + (\mu_1 + \mu_2)\min[I(U; \mathbf{Y}_{1b}), I(U; \mathbf{Y}_2)] \\
&\leq I(\mathbf{X}; \tilde{\mathbf{Y}}_{1a}|U) - I(\mathbf{X}; \mathbf{Y}_2|U) + [\mu_1 I(U; \mathbf{Y}_{1b}) + \mu_2 I(U; \mathbf{Y}_2)] \\
&= h(\mathbf{Z}_2) - h(\tilde{\mathbf{Z}}_{1a}) + \mu_1 h(\mathbf{X} + \mathbf{Z}_{1b}) + \mu_2 h(\mathbf{X} + \mathbf{Z}_2) \\
&\quad + \left[h(\mathbf{X} + \tilde{\mathbf{Z}}_{1a}|U) - \mu_1 h(\mathbf{X} + \mathbf{Z}_{1b}|U) - (\mu_2 + 1)h(\mathbf{X} + \mathbf{Z}_2|U)\right] \\
&\leq \tfrac{1}{2}\log|2\pi e \mathbf{N}_2| - \tfrac{1}{2}\log\left|2\pi e \tilde{\mathbf{N}}_1\right| + \sum_{k=1}^{2}\left[\tfrac{\mu_k}{2}\log|2\pi e(\mathbf{S} + \mathbf{N}_k)|\right] \\
&\quad + \left[h(\mathbf{X} + \tilde{\mathbf{Z}}_{1a}|U) - \mu_1 h(\mathbf{X} + \mathbf{Z}_{1b}|U) - (\mu_2 + 1)h(\mathbf{X} + \mathbf{Z}_2|U)\right] \quad (30)
\end{aligned}$$

where the last inequality follows from the facts that

$$h(\tilde{\mathbf{Z}}_{1a}) = \frac{1}{2}\log\left|2\pi e \tilde{\mathbf{N}}_1\right|,$$

$$h(\mathbf{Z}_2) = \frac{1}{2}\log|2\pi e \mathbf{N}_2|,$$

$$h(\mathbf{X} + \mathbf{Z}_{1b}) \leq \frac{1}{2}\log|2\pi e(\mathbf{S} + \mathbf{N}_1)|,$$

and

$$h(\mathbf{X} + \mathbf{Z}_2) \leq \frac{1}{2}\log|2\pi e(\mathbf{S} + \mathbf{N}_2)|.$$

Let $\mu = \mu_1 + \mu_2 + 1$ and $\lambda = \mu_1/(\mu_1 + \mu_2 + 1)$. We obtain from (25) (and Proposition 2)

$$h(\mathbf{X}+\tilde{\mathbf{Z}}_{1a}|U) - \mu_1 h(\mathbf{X} + \mathbf{Z}_{1b}|U) - (\mu_2 + 1)h(\mathbf{X} + \mathbf{Z}_2|U)$$
$$\leq \frac{1}{2}\log\left|2\pi e(\mathbf{B}^* + \tilde{\mathbf{N}}_1)\right| - \frac{\mu_1}{2}\log|2\pi e(\mathbf{B}^* + \mathbf{N}_1)| - \frac{\mu_2 + 1}{2}\log|2\pi e(\mathbf{B}^* + \mathbf{N}_2)|. \quad (31)$$

Substituting (31) into (30), we have

$$\begin{aligned}
R_1 + (\mu_1 + \mu_2)R_0 &\leq \frac{1}{2}\log|2\pi e\mathbf{N}_2| - \frac{1}{2}\log\left|2\pi e\tilde{\mathbf{N}}_1\right| + \sum_{k=1}^{2}\left[\frac{\mu_k}{2}\log|2\pi e(\mathbf{S}+\mathbf{N}_k)|\right] \\
&\quad + \left[\frac{1}{2}\log\left|2\pi e(\mathbf{B}^* + \tilde{\mathbf{N}}_1)\right| - \frac{\mu_1}{2}\log|2\pi e(\mathbf{B}^* + \mathbf{N}_1)|\right. \\
&\quad \left. -\frac{\mu_2+1}{2}\log|2\pi e(\mathbf{B}^* + \mathbf{N}_2)|\right] \\
&= \frac{1}{2}\log\left|\frac{\mathbf{B}^* + \tilde{\mathbf{N}}_1}{\tilde{\mathbf{N}}_1}\right| - \frac{1}{2}\log\left|\frac{\mathbf{B}^* + \mathbf{N}_2}{\mathbf{N}_2}\right| + \sum_{k=1}^{2}\left[\frac{\mu_k}{2}\log\left|\frac{\mathbf{S}+\mathbf{N}_k}{\mathbf{B}^* + \mathbf{N}_k}\right|\right] \\
&= \frac{1}{2}\log\left|\frac{\mathbf{B}^* + \mathbf{N}_1}{\mathbf{N}_1}\right| - \frac{1}{2}\log\left|\frac{\mathbf{B}^* + \mathbf{N}_2}{\mathbf{N}_2}\right| + \sum_{k=1}^{2}\left[\frac{\mu_k}{2}\log\left|\frac{\mathbf{S}+\mathbf{N}_k}{\mathbf{B}^* + \mathbf{N}_k}\right|\right] \quad (32)
\end{aligned}$$

for any achievable secrecy rate pair $(R_0, R_1)$ for the enhanced MIMO Gaussian broadcast channel (27). Here, the last equality follows from (24).

Finally, combining Steps 1 and 2, we conclude that any achievable secrecy rate pair for the original aligned MIMO Gaussian broadcast channel (10) is also achievable for the enhanced MIMO Gaussian broadcast channel (27). Thus, (32) holds for any achievable secrecy rate pair $(R_0, R_1)$ for the original aligned MIMO Gaussian broadcast channel (10). Since $\delta > 0$, this contradicts (20). Therefore, any achievable secrecy rate pair $(R_0, R_1)$ for the aligned MIMO Gaussian broadcast channel (10) must satisfy (11) for some $0 \preceq \mathbf{B} \preceq \mathbf{S}$. This is the desired converse result, which completes the proof of the theorem. $\square$

## 4 General MIMO Gaussian Broadcast Channel

In this section, we Theorem 1 by extending the secrecy capacity result of Theorem 4 on the aligned MIMO Gaussian broadcast channel to the general MIMO broadcast channel. As mentioned in Section 1, the achievability of the rate region (7) can be obtained from the Csiszár-Körner region (4) with proper choice of input and auxiliary variables $(U, V, \mathbf{X})$. We therefore concentrate on proving the converse part of the theorem. Also as mentioned previously, the case when both channel matrices $\mathbf{H}_1$ and $\mathbf{H}_2$ are square and invertible can be easily transformed into an aligned MIMO Gaussian broadcast channel and thus has been proved by Theorem 4. Our goal next is to *approximate* a general MIMO Gaussian broadcast channel with an aligned MIMO Gaussian broadcast channel.

Without loss of generality, we assume that the channel matrices $\mathbf{H}_1$ and $\mathbf{H}_2$ are square (but *not* necessarily invertible). If that is not the case, we can apply singular value decomposition (SVD) to show that there exists an equivalent channel that has $t \times t$ square channel matrices and the same secrecy capacity region as the original channel [9, Section V-B].

Using SVD, we can write the channel matrices as

$$\mathbf{H}_k = \mathbf{U}_k\mathbf{\Lambda}_k\mathbf{V}_k^{\mathrm{T}}, \quad k = 1, 2$$

where $\mathbf{U}_k$ and $\mathbf{V}_k$ are $t \times t$ unitary matrices, and $\mathbf{\Lambda}_k$ is diagonal. We now define a new MIMO Gaussian broadcast channel:

$$\overline{\mathbf{Y}}_k[m] = \overline{\mathbf{H}}_k \mathbf{X}[m] + \mathbf{Z}_k[m] \quad k = 1, 2 \tag{33}$$

where

$$\overline{\mathbf{H}}_k = \mathbf{U}_k(\mathbf{\Lambda}_k + \alpha \mathbf{I}_t)\mathbf{V}_k^{\mathrm{T}}$$

for some $\alpha > 0$. Note that the MIMO Gaussian broadcast channel (33) does have invertible channel matrices. By Theorem 4, the secrecy capacity, $C_s(\overline{\mathbf{H}}_1, \overline{\mathbf{H}}_2, \mathbf{S})$, under the matrix power constraint (6) is given by the set of all nonnegative rate pairs $(R_0, R_1)$ satisfying

$$\begin{aligned} R_0 &\leq \min\left(\tfrac{1}{2}\log\left|\tfrac{\overline{\mathbf{H}}_1 \mathbf{S} \overline{\mathbf{H}}_1^{\mathrm{T}} + \mathbf{I}_{r_1}}{\overline{\mathbf{H}}_1 \mathbf{B} \overline{\mathbf{H}}_1^{\mathrm{T}} + \mathbf{I}_{r_1}}\right|, \tfrac{1}{2}\log\left|\tfrac{\overline{\mathbf{H}}_2 \mathbf{S} \overline{\mathbf{H}}_2^{\mathrm{T}} + \mathbf{I}_{r_2}}{\overline{\mathbf{H}}_2 \mathbf{B} \overline{\mathbf{H}}_2^{\mathrm{T}} + \mathbf{I}_{r_2}}\right|\right) \\ R_1 &\leq \tfrac{1}{2}\log\left|\overline{\mathbf{H}}_1 \mathbf{B} \overline{\mathbf{H}}_1^{\mathrm{T}} + \mathbf{I}_{r_1}\right| - \tfrac{1}{2}\log\left|\overline{\mathbf{H}}_2 \mathbf{B} \overline{\mathbf{H}}_2^{\mathrm{T}} + \mathbf{I}_{r_2}\right| \end{aligned}$$

for some $0 \preceq \mathbf{B} \preceq \mathbf{S}$.

Further note that we can write $\mathbf{H}_k = \mathbf{D}_k \overline{\mathbf{H}}_k$ where

$$\mathbf{D}_k = \mathbf{U}_k \mathbf{\Lambda}_k (\mathbf{\Lambda}_k + \alpha \mathbf{I}_t)^{-1} \mathbf{U}_k^{\mathrm{T}}.$$

Since $\mathbf{D}_k^2 \prec \mathbf{I}_t$, we have [12, Definition 1]

$$\mathbf{X} \to \overline{\mathbf{Y}}_k \to \mathbf{Y}_k \tag{34}$$

forms a Markov chain for $k = 1, 2$. Therefore, both receivers 1 and 2 receive a better signal in the new channel (33) than in the original channel (1). Note that receiver 2 also plays the role of an eavesdropper for the confidential message $W_1$. Therefore, unlike the private message problem considered in [9], enhancing both receivers in the channel does *not* necessarily lead to an increase in the secrecy capacity region. In the following, however, we show that

$$\mathcal{C}_s(\mathbf{H}_1, \mathbf{H}_2, \mathbf{S}) \subseteq \mathcal{C}_s(\overline{\mathbf{H}}_1, \overline{\mathbf{H}}_2, \mathbf{S}) + \mathcal{O}(\mathbf{H}_2, \overline{\mathbf{H}}_2, \mathbf{S}) \tag{35}$$

where

$$\mathcal{O}(\mathbf{H}_2, \overline{\mathbf{H}}_2, \mathbf{S}) := \left\{(0, R_1) : 0 \leq R_1 \leq \frac{1}{2}\log\left|\overline{\mathbf{H}}_2 \mathbf{S} \overline{\mathbf{H}}_2^{\mathrm{T}} + \mathbf{I}_t\right| - \frac{1}{2}\log\left|\mathbf{H}_2 \mathbf{S} \mathbf{H}_2^{\mathrm{T}} + \mathbf{I}_t\right|\right\}$$

Let $(R_0, R_1)$ be an achievable secrecy rate pair for the MIMO Gaussian broadcast channel (1). By the result of Csiszár and Körner [2], there exists a collection of input and auxiliary variables $(U, V, \mathbf{X})$ satisfying the Markov relation $U \to V \to \mathbf{X}$ such that

$$\begin{aligned} R_0 &\leq \min\left[I(U; \mathbf{Y}_1), I(U; \mathbf{Y}_2)\right] \\ R_1 &\leq I(V; \mathbf{Y}_1|U) - I(V; \mathbf{Y}_2|U). \end{aligned}$$

Also by the result of Csiszár and Körner [2], the secrecy rate pair $(\overline{R}_0, \overline{R}_1)$ given by

$$\begin{aligned} \overline{R}_0 &= \min\left[I(U; \overline{\mathbf{Y}}_1), I(U; \overline{\mathbf{Y}}_2)\right] \\ \overline{R}_1 &= I(V; \overline{\mathbf{Y}}_1|U) - I(V; \overline{\mathbf{Y}}_2|U) \end{aligned}$$

is achievable for the MIMO Gaussian broadcast channel (33). By the Markov relation (34), we have

$$I(U; \mathbf{Y}_k) \leq I(U; \overline{\mathbf{Y}}_k),$$
$$I(V; \mathbf{Y}_k|U) \leq I(V; \overline{\mathbf{Y}}_k|U),$$

and

$$I(\mathbf{X}; \mathbf{Y}_k|U, V) \leq I(\mathbf{X}; \overline{\mathbf{Y}}_k|U, V)$$

for $k = 1, 2$. Hence, we have

$$R_0 - \overline{R}_0 \leq \min\left[I(U; \mathbf{Y}_1), I(U; \mathbf{Y}_2)\right] - \min\left[I(U; \overline{\mathbf{Y}}_1), I(U; \overline{\mathbf{Y}}_2)\right] \leq 0 \qquad (36)$$

and

$$\begin{aligned}
R_1 - \overline{R}_1 &\leq I(V; \mathbf{Y}_1|U) - I(V; \mathbf{Y}_2|U) - \left[I(V; \overline{\mathbf{Y}}_1|U) - I(V; \overline{\mathbf{Y}}_2|U)\right] \\
&= I(V; \overline{\mathbf{Y}}_2|U) - I(V; \mathbf{Y}_2|U) - \left[I(V; \overline{\mathbf{Y}}_1|U) - I(V; \mathbf{Y}_1|U)\right] \\
&\leq I(V; \overline{\mathbf{Y}}_2|U) - I(V; \mathbf{Y}_2|U) \\
&= I(U, V; \overline{\mathbf{Y}}_2) - I(U, V; \mathbf{Y}_2) - \left[I(U; \overline{\mathbf{Y}}_2) - I(U; \mathbf{Y}_2)\right] \\
&\leq I(U, V; \overline{\mathbf{Y}}_2) - I(U, V; \mathbf{Y}_2) \\
&= I(\mathbf{X}; \overline{\mathbf{Y}}_2) - I(\mathbf{X}; \mathbf{Y}_2) - \left[I(\mathbf{X}; \overline{\mathbf{Y}}_2|U, V) - I(\mathbf{X}; \mathbf{Y}_2|U, V)\right] \\
&\leq I(\mathbf{X}; \overline{\mathbf{Y}}_2) - I(\mathbf{X}; \mathbf{Y}_2) \\
&= I(\mathbf{X}; \overline{\mathbf{Y}}_2|\mathbf{Y}_2) \qquad (37) \\
&\leq \max_{0 \preceq \mathbf{B} \preceq \mathbf{S}} \left(\frac{1}{2}\log\left|\overline{\mathbf{H}}_2 \mathbf{B} \overline{\mathbf{H}}_2^\mathrm{T} + \mathbf{I}_t\right| - \frac{1}{2}\log\left|\mathbf{H}_2 \mathbf{B} \mathbf{H}_2^\mathrm{T} + \mathbf{I}_t\right|\right) \qquad (38) \\
&= \frac{1}{2}\log\left|\overline{\mathbf{H}}_2 \mathbf{S} \overline{\mathbf{H}}_2^\mathrm{T} + \mathbf{I}_t\right| - \frac{1}{2}\log\left|\mathbf{H}_2 \mathbf{S} \mathbf{H}_2^\mathrm{T} + \mathbf{I}_t\right| \qquad (39)
\end{aligned}$$

where (37) follows from the Markov relation (34), (38) follows from a well-known inequality due to Thomas [15, Lemma 1], and (39) follows from the fact that $\mathbf{H}_2^\mathrm{T}\mathbf{H}_2 \prec \overline{\mathbf{H}}_2^\mathrm{T}\overline{\mathbf{H}}_2$. Combining (36) and (39) established the set relationship (35).

Finally, let $\alpha \downarrow 0$ on both sides of (35). Note that $\overline{\mathbf{H}}_k \to \mathbf{H}_k$ for $k = 1, 2$, so $\mathcal{C}_s(\overline{\mathbf{H}}_1, \overline{\mathbf{H}}_2, \mathbf{S})$ converges to the rate region (7) and $\mathcal{O}(\mathbf{H}_2, \overline{\mathbf{H}}_2, \mathbf{S}) \to \{(0, 0)\}$. We thus have proved the desired converse result and completed the proof of the theorem.

## 5  Numerical Examples

In this section, we illustrate the results of Theorem 1 and Corollary 3 by numerical examples. Note that finding the boundaries of the secrecy capacity regions $\mathcal{C}_s(\mathbf{H}_1, \mathbf{H}_2, \mathbf{S})$ and $\mathcal{C}_s(\mathbf{H}_1, \mathbf{H}_2, P)$ as expressed in (7) and (9) involves solving *nonconvex* optimization programs and hence is nontrivial. Following the work in [10], we can rewrite the expressions (7) and (9) such that the optimization program for finding the boundaries of $\mathcal{C}_s(\mathbf{H}_1, \mathbf{H}_2, \mathbf{S})$ and $\mathcal{C}_s(\mathbf{H}_1, \mathbf{H}_2, P)$ become tractable for the case where each of the receivers is equipped with only one receive antenna, i.e., $r_k = 1$ for $k = 1, 2$. As we limit the discussion in this section to the single receive antenna case, the channel matrices $\mathbf{H}_k$ become the $1 \times t$ channel vectors $\mathbf{h}_k$, $k = 1, 2$.

To compute the secrecy capacity region $\mathcal{C}_s(\mathbf{h}_1, \mathbf{h}_2, \mathbf{S})$, consider re-parameterizing $(R_0, R_1)$ using $(\alpha, \gamma_0)$ as

$$\begin{aligned} R_0 &= \tfrac{1}{2}\log(1 + \alpha\gamma_0) \\ R_1 &= \tfrac{1}{2}\log(1 + \alpha(1 - \gamma_0)). \end{aligned} \quad (40)$$

Thus, to see whether a particular secrecy rate pair $(R_0, R_1)$ is inside $\mathcal{C}_s(\mathbf{h}_1, \mathbf{h}_2, \mathbf{S})$ as expressed in (7), one may check, instead, whether there exists a positive semidefinite matrix $\mathbf{B}$ which satisfies the set of constraints:

$$\begin{aligned} \mathbf{h}_1(\mathbf{S} - \mathbf{B})\mathbf{h}_1^T &\geq \alpha\gamma_0(\mathbf{h}_1\mathbf{B}\mathbf{h}_1^T + 1) \\ \mathbf{h}_2(\mathbf{S} - \mathbf{B})\mathbf{h}_2^T &\geq \alpha\gamma_0(\mathbf{h}_2\mathbf{B}\mathbf{h}_2^T + 1) \\ \mathbf{h}_1\mathbf{B}\mathbf{h}_1^T - \mathbf{h}_2\mathbf{B}\mathbf{h}_2^T &\geq \alpha(1 - \gamma_0)(\mathbf{h}_2\mathbf{B}\mathbf{h}_2^T + 1) \\ \mathbf{B} &\preceq \mathbf{S}. \end{aligned} \quad (41)$$

Note that all the constraints in (41) are linear in $\mathbf{B}$. Hence, whether there exists a feasible solution can be examined using standard semidefinite programming techniques (i.e., CVX, a package for specifying and solving convex programs [16]). Note from (40) that both $R_0$ and $R_1$ increase as $\alpha$ increases. Therefore, for a fixed $\gamma_0$, a boundary point of $\mathcal{C}_s(\mathbf{h}_1, \mathbf{h}_2, \mathbf{S})$ can be found by searching over the maximum $\alpha$ such that the set of constraints in (41) admits a feasible solution. Sweeping over $\gamma_0 \in [0, 1]$ gives all the boundary points of $\mathcal{C}_s(\mathbf{h}_1, \mathbf{h}_2, \mathbf{S})$.

Similarly, to compute the secrecy capacity region $\mathcal{C}_s(\mathbf{h}_1, \mathbf{h}_2, P)$, we consider the set of constraints for a pair of positive semidefinite matrices $(\mathbf{B}_1, \mathbf{B}_2)$:

$$\begin{aligned} \mathbf{h}_1\mathbf{B}_2\mathbf{h}_1^T &\geq \alpha\gamma_0(\mathbf{h}_1\mathbf{B}_1\mathbf{h}_1^T + 1) \\ \mathbf{h}_2\mathbf{B}_2\mathbf{h}_2^T &\geq \alpha\gamma_0(\mathbf{h}_2\mathbf{B}_1\mathbf{h}_2^T + 1) \\ \mathbf{h}_1\mathbf{B}_1\mathbf{h}_1^T - \mathbf{h}_2\mathbf{B}_1\mathbf{h}_2^T &\geq \alpha(1 - \gamma_0)(\mathbf{h}_2\mathbf{B}_1\mathbf{h}_2^T + 1) \\ \mathrm{Tr}(\mathbf{B}_1 + \mathbf{B}_2) &\leq P. \end{aligned} \quad (42)$$

Again, all the constraints in (42) are linear in $(\mathbf{B}_1, \mathbf{B}_2)$ so whether there exists a feasible solution can be examined using standard semidefinite programming techniques [16]. Therefore, for a fixed $\gamma_0$, a boundary point of $\mathcal{C}_s(\mathbf{h}_1, \mathbf{h}_2, P)$ can be found by searching over the maximum $\alpha$ such that the set of constraints in (42) admits a feasible solution. Sweeping over $\gamma_0 \in [0, 1]$ gives all the boundary points of $\mathcal{C}_s(\mathbf{h}_1, \mathbf{h}_2, P)$.

Figure 3 plots the secrecy capacity regions $\mathcal{C}_s(\mathbf{h}_1, \mathbf{h}_2, \mathbf{S})$ and $\mathcal{C}_s(\mathbf{h}_1, \mathbf{h}_2, P)$ for the channel vectors $\mathbf{h}_1 = [2\ 0.4]$ and $\mathbf{h}_2 = [0.4\ 1]$ and power constraints

$$\mathbf{S} = \begin{bmatrix} 3.3333 & 1.2346 \\ 1.2346 & 1.6667 \end{bmatrix}$$

and $P = \mathrm{Tr}(\mathbf{S}) = 5$. For comparison, in Figure 3, we have also plotted the capacity regions of the same MIMO Gaussian broadcast channel with a common message $W_0$ intended for both receiver 1 and 2 and a private message $W_1$ intended only for receiver 1 (but without any

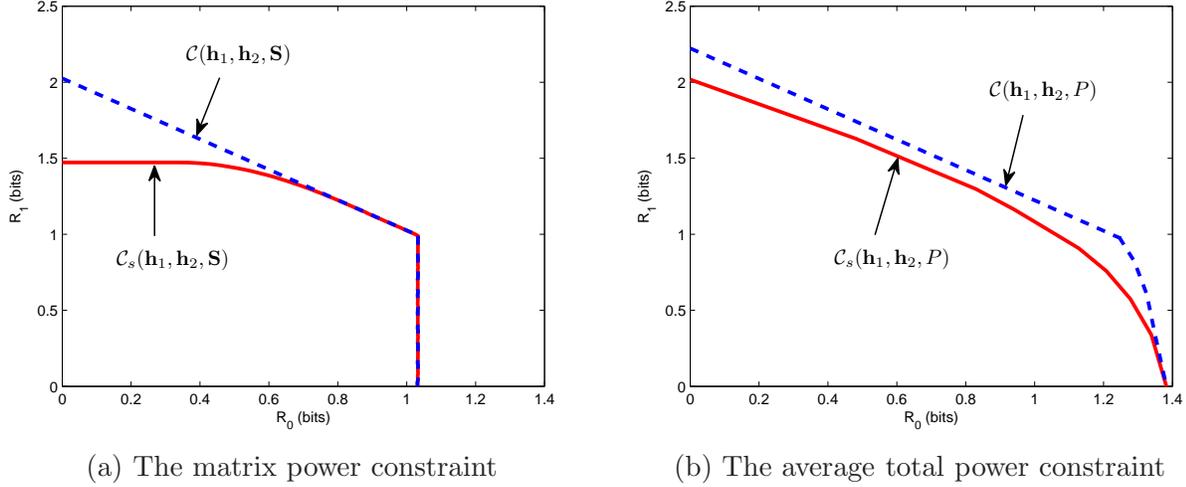

(a) The matrix power constraint  (b) The average total power constraint

Figure 3: An illustration of the secrecy capacity regions of the MIMO Gaussian broadcast channel with common and confidential messages.

secrecy constraints). This problem is known as the MIMO Gaussian broadcast channel with degraded message sets [10, 17]. As shown in [10], the capacity region, $\mathcal{C}(\mathbf{h}_1, \mathbf{h}_2, \mathbf{S})$, under the matrix power constraint (6) is given by[2]

$$\mathcal{C}(\mathbf{h}_1, \mathbf{h}_2, \mathbf{S}) = \mathcal{R}_1(\mathbf{h}_1, \mathbf{S}) \cap \mathcal{R}_2(\mathbf{h}_1, \mathbf{h}_2, \mathbf{S})$$

where $\mathcal{R}_1(\mathbf{h}_1, \mathbf{S})$ is given by the nonnegative rate pairs $(R_0, R_1)$ satisfying

$$R_0 + R_1 \leq \frac{1}{2} \log(\mathbf{h}_1 \mathbf{S} \mathbf{h}_1^\mathrm{T} + 1),$$

and $\mathcal{R}_2(\mathbf{h}_1, \mathbf{h}_2, \mathbf{S})$ is given by the nonnegative rate pairs $(R_0, R_1)$ satisfying

$$\begin{array}{rcl} R_0 & \leq & \frac{1}{2} \log\left( \frac{\mathbf{h}_2 \mathbf{S} \mathbf{h}_2^\mathrm{T} + 1}{\mathbf{h}_2 \mathbf{B} \mathbf{h}_2^\mathrm{T} + 1} \right) \\ R_1 & \leq & \frac{1}{2} \log(\mathbf{h}_1 \mathbf{B} \mathbf{h}_1^\mathrm{T} + 1) \end{array}$$

for some $0 \preceq \mathbf{B} \preceq \mathbf{S}$. Similarly, the capacity region, $\mathcal{C}(\mathbf{h}_1, \mathbf{h}_2, P)$, under the average total power constraint (2) is given by

$$\mathcal{C}(\mathbf{h}_1, \mathbf{h}_2, P) = \mathcal{R}_1(\mathbf{h}_1, P) \cap \mathcal{R}_2(\mathbf{h}_1, \mathbf{h}_2, P)$$

where $\mathcal{R}_1(\mathbf{h}_1, P)$ is given by the nonnegative rate pairs $(R_0, R_1)$ satisfying

$$R_0 + R_1 \leq \frac{1}{2} \log(P \|\mathbf{h}_1\|^2 + 1),$$

---

[2] As shown in [10], this result holds for the general MIMO Gaussian broadcast channel, not just for the single receive antenna case.

and $\mathcal{R}_2(\mathbf{h}_1, \mathbf{h}_2, P)$ is given by the nonnegative rate pairs $(R_0, R_1)$ satisfying

$$\begin{aligned} R_0 &\leq \tfrac{1}{2} \log \left( \tfrac{\mathbf{h}_2(\mathbf{B}_1+\mathbf{B}_2)\mathbf{h}_2^T+1}{\mathbf{h}_2 \mathbf{B}_1 \mathbf{h}_2^T+1} \right) \\ R_1 &\leq \tfrac{1}{2} \log(\mathbf{h}_1 \mathbf{B}_1 \mathbf{h}_1^T + 1) \end{aligned}$$

for some $\mathbf{B}_1 \succeq 0$, $\mathbf{B}_2 \succeq 0$ and $\mathrm{Tr}(\mathbf{B}_1 + \mathbf{B}_2) \leq P$. The boundaries of the rate regions $\mathcal{R}_2(\mathbf{h}_1, \mathbf{h}_2, \mathbf{S})$ and $\mathcal{R}_2(\mathbf{h}_1, \mathbf{h}_2, P)$ can be computed similarly to those of $\mathcal{C}_s(\mathbf{h}_1, \mathbf{h}_2, \mathbf{S})$ and $\mathcal{C}_s(\mathbf{h}_1, \mathbf{h}_2, P)$, respectively. As expected, for any given common rate $R_0$, the maximum secrecy rate is less than (or equal to) the maximum private rate due to the additional secrecy constraint at receiver 2.

## A  Proof of the Converse Part of Proposition 1

In this proof, we use $X_i^j$ to denote the vector $(X[i], X[i+1], \ldots, X[j])$, and when $i = 1$, we further simplify the notation by using $X^j$ to denote the vector $(X[1], X[2], \ldots, X[j])$. We also use $X_i$ to denote $X[i]$.

We consider a $(2^{nR_0}, 2^{nR_1}, n)$ code with the average block error probability $P_e^{(n)}$. Then we have the following joint probability distribution

$$p(w_0, w_1, x^n, \tilde{y}_{1a}^n, y_{1b}^n, y_2^n) = p(w_0)p(w_1)p(x^n|w_0 w_1) \prod_{i=1}^{n} [p(\tilde{y}_{1ai}|x_i)p(y_{1bi}y_{2i}|\tilde{y}_{1ai})]. \qquad (43)$$

By Fano's inequality, we have

$$H(W_0|Y_{1b}^n) \leq nR_0 P_e^{(n)} + 1 := n\delta_{1n} \qquad (44)$$
$$H(W_0|Y_2^n) \leq nR_0 P_e^{(n)} + 1 := n\delta_{1n} \qquad (45)$$
$$H(W_1|\tilde{Y}_{1a}^n) \leq nR_1 P_e^{(n)} + 1 := n\delta_{2n} \qquad (46)$$

where $\delta_{1n}, \delta_{2n} \to 0$ if $P_e^{(n)} \to 0$.

We define the following auxiliary random variable:

$$U_i := (W_0, \tilde{Y}_{1a}^{i-1}) \qquad (47)$$

which satisfies the Markov chain relationship

$$U_i \to X_i \to (\tilde{Y}_{1ai}, Y_{1bi}, Y_{2i}).$$

We first bound $R_0$ as follows.

$$nR_0 = H(W_0) \leq I(W_0; Y_{1b}^n) + n\delta_{1n} \quad (48)$$

$$= \sum_{i=1}^{n} I(W_0; Y_{1bi}|Y_{1b}^{i-1}) + n\delta_{1n}$$

$$\leq \sum_{i=1}^{n} I(W_0, \tilde{Y}_{1a}^{i-1}; Y_{1bi}|Y_{1b}^{i-1}) + n\delta_{1n}$$

$$\leq \sum_{i=1}^{n} I(W_0, \tilde{Y}_{1a}^{i-1}, Y_{1b}^{i-1}; Y_{1bi}) + n\delta_{1n}$$

$$\leq \sum_{i=1}^{n} I(W_0, \tilde{Y}_{1a}^{i-1}; Y_{1bi}) + n\delta_{1n} \quad (49)$$

$$\leq \sum_{i=1}^{n} I(U_i; Y_{1bi}) + n\delta_{1n} \quad (50)$$

where (48) follows from Fano's inequality (44), and (49) follows from the degradedness condition, i.e., $(W_0, Y_{1bi}) \to \tilde{Y}_{1a}^{i-1} \to Y_{1b}^{i-1}$. We can follow the steps similar to those in (48)-(50) with $Y_{1b}$ being replaced by $Y_2$, and obtain the following bound

$$nR_0 \leq \sum_{i=1}^{n} I(U_i; Y_{2i}) + n\delta_{1n}. \quad (51)$$

We now bound $nR_1$ and obtain

$$nR_1 = H(W_1|Y_2^n) \tag{52}$$
$$= H(W_1|W_0, Y_2^n) + I(W_1; W_0|Y_2^n)$$
$$\leq H(W_1|W_0, Y_2^n) + n\delta_{1n}$$
$$= I(W_1; \tilde{Y}_{1a}^n|W_0, Y_2^n) + H(W_1|W_0, Y_2^n, \tilde{Y}_{1a}^n) + n\delta_{1n}$$
$$\leq I(W_1; \tilde{Y}_{1a}^n|W_0, Y_2^n) + n\delta_{2n} + n\delta_{1n} \tag{53}$$
$$\leq I(W_1, X^n; \tilde{Y}_{1a}^n|W_0, Y_2^n) + n\delta_{2n} + n\delta_{1n}$$
$$= I(X^n; \tilde{Y}_{1a}^n|W_0, Y_2^n) + n\delta_{2n} + n\delta_{1n} \tag{54}$$
$$= H(X^n|W_0, Y_2^n) - H(X^n|W_0, Y_2^n, \tilde{Y}_{1a}^n) + n\delta_{2n} + n\delta_{1n}$$
$$= H(X^n|W_0, Y_2^n) - H(X^n|W_0, \tilde{Y}_{1a}^n) + n\delta_{2n} + n\delta_{1n} \tag{55}$$
$$= I(X^n; \tilde{Y}_{1a}^n|W_0) - I(X^n; Y_2^n|W_0) + n\delta_{2n} + n\delta_{1n}$$
$$= \sum_{i=1}^{n} \left[ I(X^n; \tilde{Y}_{1ai}|\tilde{Y}_{1a}^{i-1}, W_0) - I(X^n; Y_{2i}|Y_2^{i-1}, W_0) \right] + n\delta_{2n} + n\delta_{1n}$$
$$= \sum_{i=1}^{n} \Big[ H(\tilde{Y}_{1ai}|\tilde{Y}_{1a}^{i-1}, W_0) - H(\tilde{Y}_{1ai}|\tilde{Y}_{1a}^{i-1}, W_0, X^n)$$
$$\qquad - H(Y_{2i}|Y_2^{i-1}, W_0) + H(Y_{2i}|Y_2^{i-1}, W_0, X^n) \Big] + n\delta_{2n} + n\delta_{1n}$$
$$\leq \sum_{i=1}^{n} \Big[ H(\tilde{Y}_{1ai}|\tilde{Y}_{1a}^{i-1}, W_0) - H(\tilde{Y}_{1ai}|\tilde{Y}_{1a}^{i-1}, W_0, X_i)$$
$$\qquad - H(Y_{2i}|\tilde{Y}_{1a}^{i-1}, Y_2^{i-1}, W_0) + H(Y_{2i}|Y_2^{i-1}, W_0, X_i) \Big] + n\delta_{2n} + n\delta_{1n} \tag{56}$$
$$\leq \sum_{i=1}^{n} \Big[ H(\tilde{Y}_{1ai}|\tilde{Y}_{1a}^{i-1}, W_0) - H(\tilde{Y}_{1ai}|\tilde{Y}_{1a}^{i-1}, W_0, X_i)$$
$$\qquad - H(Y_{2i}|\tilde{Y}_{1a}^{i-1}, W_0) + H(Y_{2i}|\tilde{Y}_{1a}^{i-1}, W_0, X_i) \Big] + n\delta_{2n} + n\delta_{1n} \tag{57}$$
$$= \sum_{i=1}^{n} \left[ I(X_i; \tilde{Y}_{1ai}|U_i) - I(X_i; Y_{2i}|U_i) \right] + n\delta_{2n} + n\delta_{1n} \tag{58}$$

where (52) follows from perfect secrecy condition, (53) follows from Fano's inequality, (54) follows from the Markov chain $(W_0, W_1) \to (X^n, Y_2^n) \to \tilde{Y}_{1a}^n$, (55) follows from the degradedness condition, i.e., $(X^n, W_0) \to \tilde{Y}_{1a}^n \to Y_2^n$, (56) follows from the Markov chain relationship $(\tilde{Y}_{1a}^{i-1}, W_0, X^n) \to X_i \to \tilde{Y}_{1ai}$ and conditioning does not increase entropy, and (57) follows from the Markov chain relationships $(Y_{2i}, W_0) \to \tilde{Y}_{1a}^{i-1} \to Y_2^{i-1}$ and $(Y_2^{i-1}, \tilde{Y}_{1a}^{i-1}, W_0) \to X_i \to Y_{2i}$.

The single-letter outer bound can be obtained by letting $J$ be a time-sharing variable uniformly distributed over $\{1, \ldots, n\}$, and define $U = (U_J, J)$, $X = X_J$, $\tilde{Y}_{1a} = \tilde{Y}_{1aJ}$, $Y_{1b} = Y_{1bJ}$, and $Y_2 = Y_{2J}$.